\documentclass[12pt]{article}

\usepackage{amsmath,amssymb,latexsym}

\def\hybrid{\topmargin -20pt  \oddsidemargin 0pt
         \headheight 0pt   \headsep 0pt
         \textwidth 6.25in 
         \textheight 9.5in 
         \marginparwidth .875in
         \parskip 5pt plus 1pt   \jot = 1.5ex}

\hybrid

\def\o+{\oplus}
\def\ra{\rightarrow}
\def\Lra{\Longrightarrow}
\def\beqa{\begin{eqnarray}}
\def\eeqa{\end{eqnarray}}
\def\del{\partial}
\def\al{\alpha}
\def\ga{\gamma}
\def\de{\delta}

\def\si{\sigma}

\def\la{\lambda}
\def\om{\omega}
\def\De{\Delta}
\def\La{\Lambda}
\def\Om{\Omega}
\def\G{{\cal G}}
\def\O{{\cal O}}
\def\T{{\cal T}}
\def\tr{\text{tr}}
\def\Re{{\text{Re}}}
\def\Im{{\text{Im}}}

\newcommand{\resetcounter}{\setcounter{equation}{0}}

\begin{document}

\thispagestyle{empty}
\rightline{LMU-ASC 05/05, MPP-2005-13, UMD-PP-05-039}
\rightline{hep-th/0502168}
\vspace{1truecm} \centerline{\bf \LARGE Moduli Stabilization}
\vspace{0.3truecm} \centerline{\bf \LARGE in the Heterotic/IIB
Discretuum}

\vspace{.6truecm} \centerline{Gottfried
Curio$^{a,}$\footnote{curio@theorie.physik.uni-muenchen.de}, Axel
Krause$^{b,c,}$\footnote{krause@physics.umd.edu} and Dieter
L\"ust$^{a,d,}$\footnote{luest@mppmu.mpg.de,
luest@theorie.physik.uni-muenchen.de}}

\vspace{.5truecm}

\centerline{{\em $^a$Arnold-Sommerfeld-Center for Theoretical
Physics}} \centerline{{\em Department f\"ur Physik,
Ludwig-Maximilians-Universit\"at M\"unchen}} \centerline{{\em
Theresienstra\ss e 37, 80333 M\"unchen, Germany}}

\vspace{.2truecm}

{\em \centerline{$^b$Department of Physics, University of
Maryland, College Park, MD 20742, USA}}

\vspace{.2truecm}

{\em \centerline{$^c$Perimeter Institute for Theoretical Physics,
Waterloo, Ontario N2L 2Y5, Canada}}

\vspace{.2truecm}

{\em \centerline{$^d$Max-Planck-Institut f\"ur Physik, F\"ohringer
Ring 6, 80805 M\"unchen, Germany}}



\begin{abstract}
We consider supersymmetric compactifications of type IIB and
weakly coupled heterotic string-theory in the presence of $G$
resp.~$H$-flux and various non-perturbative effects. We point out
that non-perturbative effects change the Hodge structure of the
allowed fluxes in type IIB significantly. In the heterotic case it
is known that, in contrast to the potential read off from
dimensional reduction, the effective four-dimensional description
demands for consistency a non-vanishing $H^{2,1}$ component, once
a non-trivial $H^{3,0}$ component balances the gaugino condensate.
The $H^{2,1}$ causes classically (but not when non-perturbative
effects are included) a non-K\"ahler compactification geometry
whose moduli space is, however, poorly understood. We show that
the occurrence of $H^{2,1}$ could be avoided with world-sheet
instantons by using a KKLT-like two-step procedure for moduli
stabilization. Moreover, heterotic moduli stabilization under the
inclusion of one-loop corrections to the gauge kinetic function
led to negative gauge couplings and a corresponding strong
coupling transition. This problem disappears, as well, when
world-sheet instantons are included. They stabilize moreover the
K\"ahler modulus without the need for a non-K\"ahler geometry with
non-trivial $dJ$.
\end{abstract}

\vspace{-5mm}


\newpage
\pagenumbering{arabic}

\section {Introduction and Summary}

In this paper we will investigate the problem of moduli
stabilization in type IIB and heterotic string supersymmetric flux
compactifications with additional contributions to the effective
superpotentials, most notably gaugino condensation or also wrapped
Euclidean D3-branes. We will see that the discrete landscape
considerably differs, notably in the Hodge types of the allowed
flux, from those groundstates obtained from a pure 3-form flux
superpotential.

On the type IIB side the analysis is to very large extent
motivated by the KKLT scenario [\ref{KKLT}] where the 3-form flux
superpotential is augmented by additional terms which depend on a
size modulus
of the compact internal space $X$. On the heterotic side we will
study the combined effect of an effective superpotential which
contains the NS 3-form flux $H$, the gaugino condensate,
world-sheet instanton effects and possibly terms that describe the
deviation from $X$ being K\"ahler. Both the heterotic and the type
IIB case exhibit several similarities and analogies, which should
be explainable from some underlying string-string duality
symmetry. We will exhibit these analogies and discuss them in the
context of the effective actions, but we will not give a serious
attempt to trace them back to some string-string duality
transformations (which should nevertheless be possible in some
explicit orientifold/heterotic dual pairs). In both cases the
result of the discussion will be rather similar: the inclusion of
the additional effects in the superpotential besides the 3-form
fluxes has the effect that generic supersymmetric groundstates are
described in the type IIB case by fluxes which are not anymore ISD
(imaginary self-dual) with only $G^{2,1}$ and $G^{0,3}$ components
but rather will include all IASD (imaginary anti self-dual) types
as well, respectively in heterotic compactifications there will
generically be $H$-fluxes of type (2,1) and (1,2) besides the
`usual' (3,0) $H$-flux. We show the necessity of the more general
Hodge type by arguing that $H^{2,1}=0$ is generically impossible
to impose consistently. Only in a kind of two step procedure
(which constitutes an additional assumption), applied in the
original work of KKLT, where one first fixes the complex structure
moduli, and then solves the supersymmetry conditions for the
remaining fields, the original Hodge structure for $G$-
resp.~$H$-flux can be preserved.

{\em Superpotentials and moduli stabilization in heterotic
compactifications}

The procedure used in the heterotic string theory bears a number
of interesting similarities and important differences in
comparison with the type IIB situation. Firstly the complex
structure moduli $z_i$ are now fixed by enforcing a
proportionality between $H^{3,0}$ and $\Om$; here the first is
being assumed to get contributions just from the $dB$ sector (as
is the case in the standard embedding). The quantization
[\ref{Rohm Witten}] of $H=dB$ leads then to a corresponding fixing
of the periods. The needed proportionality stems from a balancing
between the gaugino condensate in the hidden sector and the
$H^{3,0}+ c.c.$ flux. This, and some details of this situation
that are recalled below in subsect.~\ref{3 flux het}, comes from
the corresponding complete square in the heterotic string
Lagrangian. While this mechanism alone, realising a no scale
scenario, will break supersymmetry and leave the $T$-modulus
unfixed with  a still vanishing (tree-level) potential, it can be
promoted to a $T$-stabilizing supersymmetric $AdS$ model by
including size-dependent effects like one-loop gauge coupling
corrections.

It was shown [\ref{BdA}] that in heterotic compactifications it is
not possible to turn on consistently (in a supersymmetric way, say
in the $z_i$ and $S$ sectors) the flux component $H^{3,0}+c.c.$
{\em alone}. A component $H^{2,1}+c.c.$ will be induced
automatically (even if it is only a small quantum effect). Having
such a component in the heterotic string implies [\ref{S}]
{\em classically} that the underlying
compactification manifold cannot be K\"ahler any longer if
supersymmetry is to be preserved (some further details from more
recent investigations on this set-up will be recalled below, too).
But in that case it is not quite clear what the appropriate moduli
replacing or generalising the complex structure moduli and the
K\"ahler moduli will be;
therefore one would have, taking this seriously, a somewhat
inconsistent starting point when using superpotentials for the
ordinary moduli. Hence one would prefer to avoid the occurrence of
the $dJ$ component.

A strong argument in favor of having a non-trivial $dJ\neq 0$ has
been its stabilization effect on the overall radial modulus $T$
[\ref{S}], [\ref{CCdAL2}], [\ref{B1}]. In this work we will show
in section \ref{WSI sect} that by adding the size-dependent
world-sheet instanton effect
\beqa
W_{WSI}=Be^{-bT}
\eeqa
to build the combined superpotential
\beqa
\label{W full}
W=W_H[z_i] + W_{WSI}[T] + W_{GC}[S] \; ,
\eeqa
one can stabilize all moduli supersymmetrically, including the
radial modulus $T$, without the need for a non-trivial $dJ$. In
view of the unknown moduli space of non-K\"ahler spaces this is an
interesting result which allows to carry out the moduli
stabilization program rigorously within the currently known
mathematical framework.

We first discuss this in the scenario where the complex structure
moduli do not acquire masses above the mass scale for the
stabilized $T$-modulus and therefore do not decouple from its
dynamics. Here we point out that when solving the supersymmetry
conditions for all moduli with vanishing $H^{2,1}$ one generically
overconstrains the $H$-field. More precisely, when one wants to
allow for a potential solubility of the constraints even in
principle, it is crucial to take into account the non-trivial
dependence of the one-loop determinant $B$ on the complex
structure moduli $z_i$
\beqa
\label{Pfaff}
B = \frac{\mbox{Pfaff}\, \bar{\del}_{V(-1)}|_C}
{(\mbox{det}\bar{\del}_{{\cal O}(-1)}|_C)^2} \; .
\eeqa
(here a constant is suppressed and actually for reasons of
well-definedness the exponential $T$ factor should be included,
cf.~[\ref{W}]). $C$ is a contributing genus zero curve which is
assumed to be isolated so that its normal bundle\footnote{The
denominator of (\ref{Pfaff}) is $\mbox{det}\, \del_{y_i y_j}W_2$
with the superpotential $W_2(C = \del D)=\int_D \Om$ (up to
additive constants) and $y_i$ local coordinates of $N$. $W_2$ is
stationary for holomorphic $C$ [\ref{BW}], cf.~[\ref{W QCD}].} is
$N={\cal O}(-1)\oplus {\cal O}(-1)$ (we will not consider the
dependence of $B$ on vector bundle moduli in this paper). That is,
it is important to note that
\beqa
\del_{z_i} \, B \neq 0 \; ,
\eeqa
which violates the seemingly decoupled structure of the three
contributions in (\ref{W full}) with respect to the $z_i$, $T$ and
$S$ dependence. But, since a suitable adjustment of $H$ can
nevertheless not be ensured generically, we also discuss the
two-step procedure as an alternative.

{\em Comparison with moduli stabilization in type IIB}

Recall that in [\ref{Nilles}] effective supergravity descriptions
for type IIB with $D3$ and/or $D7$ branes along the lines of
[\ref{KKLT}] and [\ref{BKQ}] resp.~a heterotic theory with flux
and gaugino condensation were considered in parallel. On the type
IIB side this amounts to the consideration of the superpotential
\beqa
W=A+B\tau +Ce^{-aT}\;\;\;\;\;\;\;\; \mbox{with} \;\; A=\int
F\wedge \Om \;\;, \;\; B=-\int H\wedge \Om
\eeqa
or, when the dependence of $A$ and $B$ on the complex structure
moduli $z_i$ is considered, to the investigation of
$W=W_{eff}^{\tau}[\tau] + C e^{-aT}$ after the $z_i$ have been
integrated out.

On the heterotic side, after the replacement $\tau \ra T$ and
$T\ra S$ the superpotential is
\beqa
W=A+BT + C e^{-aS}\;\;\;\;\;\;\;\; \mbox{with} \;\; A=W_H = \int H
\wedge \Om \;\; , \;\; B =-{i\over 2}\int dJ\wedge \Omega\, .
\eeqa
We expect that inclusion of the $dJ$ completion of $H$ makes for a
full analogy to the type IIB situation in many respects. In the
following we do not invoke the term $\frac{i}{2}\int \Om \wedge
dJ$ so that we have $B=0$. The omission of this term, which keeps
$H$ purely real, will be the reason for the impossibility to
stabilize a complex structure modulus $z$ exponentially near to a
conifold vacuum $z=0$, cf.~the remarks after (\ref{z equ}).
Further one has the problem of weak coupling stabilization of the
heterotic dilaton. To draw the parallel to the type IIB procedure,
when speaking next about complex structure moduli $z_i$, we
restrict (having $B=0$) to the K\"ahler case as otherwise a
rationale for grouping the moduli in `K\"ahler moduli' and
`complex structure moduli' (both may be in some generalized sense)
is not well understood; pragmatically we consider (here for the
case without the one-loop corrections) with [\ref{Nilles}] a
decomposition of $W$ where one has $W=W_{eff}^T[T] + W_{GC}$, i.e.
\beqa
\label{W eff}
W_{eff}^T[T] = W_H + Be^{-bT}
\eeqa

Using this `substitution'-dictionary $T \leftrightarrow S$ between
type IIB and the heterotic theory one finds that a number of
problems of the heterotic theory can be addressed successfully,
partly when the two-step procedure is employed.

Firstly, the pertinent problem of stabilization of the heterotic
dilaton at weak coupling (large $S_R$) mirrors the analogous
problem of the consistency of the SUGRA approximation in the type
IIB investigation: there too it was found that at the minimum (or
rather: stationary point) of the superpotential $W=W_{flux} +
W_{non-pert}$ (with $W_{non-pert} = Ce^{-aT}$ or $Ce^{-aS}$ in
type IIB and the heterotic theory, respectively) the two
contributions have to balance each other approximately. This leads
to the analogous problem as one encounters in the heterotic
dilaton stabilization: for consistency of the analysis one needs
large $T_R$ (resp. $S_R$) but the flux is integral and the period
generically of order 1. It is then the degree of freedom which
stems from having many 3-cycles which comes to the rescue in the
type IIB case thereby making possible a exponentially small
$W_{flux}$. There are important differences to this in the
heterotic case (cf. discussion in subsect. 3.6). It is also
interesting to note that if one would try to make $W_{flux}$ small
by having just one flux on a conifold cycle and tries to make the
corresponding $z$-period small as in [\ref{GKP}] one encounters
the difficulty that the heterotic 3-form is {\em real} and
therefore the near-conifold vacuum ($z$ exponentially close to
zero) can not be stabilized.

Secondly adopting the two-step procedure of KKLT in the heterotic
case by fixing the $z_i$ first by using just the $W_{flux}$ alone,
one can avoid the occurrence of a $H^{2,1}$ component. In the
paper we will write down the $D_i W=0$ conditions first in the
full form and then point to the emerging $H^{2,1}$ resp. its
avoidance indicated here. In this case we find that the remaining
dilaton and radial moduli, $S$ and $T$, can both be stabilized
reliably without the need for a non-trivial $dJ$.

Here we spent some time to show that the greater flexibility in
the heterotic string of having not just $H=dB$ will not be enough
to avoid $H^{2,1}$ independently of the use of the two-step
procedure. One might have guessed that this could be possible as
the relevant condition that $H=dB -(CS_{YM} - CS_L)$ has type $3,0
+ c.c.$ does not lead immediately to a second set of conditions on
the periods (as in [\ref{Rohm Witten}]) as the $CS$ sector is not
quantized (the usual argument for $z_i$ stabilization relies on
$dB$, and then therefore also a multiple of $\Om + c.c.$, being
integrally quantized [\ref{Rohm Witten}]). But this sector is
difficult to control (as it comes naturally with its own
supersymmetry conditions, cf.~below) if $A_{YM}$ is not just flat
(cf.~[\ref{GKLM}]) in which case $CS_{YM}$ is again quantized.

{\em Avoiding the strong coupling transition problem}

Furthermore, and independently of the $H^{2,1}$ issue, we will try
to solve the strong coupling transition puzzle which arose in
[\ref{GKLM}] where vacua based on $H^{3,0}$ flux and gaugino
condensation contributions were considered. It was found there
that one-loop corrections $f_{obs/hid} = S \mp \beta T$ to the
gauge couplings led to negative gauge couplings
\beqa
\label{transition problem} \Re f_{obs} < 0 \; .
\eeqa
This would have implied a not well understood strong coupling
transition. We will show that once world-sheet instantons are
additionally taken into account this problem disappears. Likewise
the unorthodox choice $\Re f_{obs} < \Re f_{hid}$, which was found
to be necessary in [\ref{GKLM}], will no longer be needed.

We finally prove the stability of the vacuum along the lines of
[\ref{Nilles}] by checking the stability criterion of
[\ref{Nilles}] actually for the relevant $W_{eff}^T[T]$.

The paper is structured in the following way. In the next section
we will discuss the supersymmetric ground states in type IIB with
3-form fluxes as well as with additional contributions to the
superpotential and point to the occurrence of IASD flux components
even in the supersymmetric case. Then we will introduce the
heterotic superpotential and exploit several analogies with the
type IIB superpotential. We will discuss the possibility of having
just an $H^{3,0}+c.c.$-flux balancing the gaugino condensate and
show that it is only possible to avoid the occurrence of a further
$H^{2,1}+c.c.$-flux by implementing a KKLT-like two-step procedure
also into the heterotic compactifications. Note that $dJ\neq 0$ is
no longer simply enforced by $H^{2,1}\neq 0$ (as in the classical
case $D_i W_{flux}=0$) when non-perturbative effects are included
which now (instead of the previous $dJ$) balance an $H^{2,1}$. One
positive feature of having a $dJ$ component was that it would fix
the size of the overall radial modulus we proceed to discuss an
alternative mechanism to achieve this. We will investigate the
inclusion of a non-perturbative size-fixing heterotic
superpotential from world-sheet instantons. In this framework we
can solve furthermore the strong coupling transition problem which
was present in $T$-modulus stabilization from including one-loop
corrections to the gauge kinetic functions. We finally check the
stability of the vacuum along the lines of [\ref{Nilles}].

\section{Moduli Stabilization in Type IIB with 3-Form Fluxes
and Non-Perturbative Corrections}

\resetcounter

Let us now consider the type IIB case and discuss which are the
conditions on the 3-form flux $G$ for supersymmetric vacua taking
into account K\"ahler moduli\ dependent corrections to the flux
superpotential, cf.~[\ref{KKLT}], [\ref{Nilles}]. Since there is a
striking analogy between the type IIB and the heterotic moduli
stabilization procedures, we will be particularly interested in
the question under which conditions the $G^{1,2}$ flux component
arises; this will help us to better understand later the
occurrence of the heterotic $H^{2,1}+c.c.$ component. Specifically
we are considering the following type IIB superpotential
\beqa\label{typeiiBsupo}
W = W_G + W_{D3I} +\Big[ W_S \Big]
= \int G\wedge \Om + C(z_i)\, e^{-aT} \;\;\;\; \Big[ + C'(z_i)\, e^{-bS} \Big]
\eeqa
Here $S=s+i\si_S$ denotes  $-i\tau$ with $\tau$ the type IIB
dilaton so that $G=F-\tau H = F - iSH$. $T$ denotes\footnote{often
called $\rho$ in the literature; for the comparison with the
heterotic string we switch the notation, cf. also [\ref{Nilles}]},
by abuse of language, the four-cycle volume of the $D3$ instanton.
It is closely related to the proper K\"ahler modulus measuring a
two-cycle volume. This difference will however not being important
here.

The first term $W_G$ is the standard type IIB flux superpotential
\beqa
W_G=A+B\tau \;\;\;\;\;\;\;\; \mbox{with} \;\; A=\int F\wedge \Om
\;\;, \;\; B=-\int H\wedge \Om
\eeqa
The supersymmetric vacua which follow from the flux superponetial
are obtained if the flux is ISD and and of the form $G^{2,1}$
[\ref{GKP}] (more discussion on flux vacua can be found e.g.
[\ref{fluxes}]).

The second term $W_{D3I}$ is the correction due to Euclidean
D3-branes wrapped around 4-cycles in the $X$. For them to
contribute to the superpotential the four-fold used for F-theory
compactification has to admit divisors of arithmetic genus one,
which project to 4-cycles in the base $X$. Alternatively,
$W_{D3I}$ can originate from non-perturbative gaugino condensation
in some hidden, asymptotically free gauge group. The difference
between wrapped Euclidean D3-branes and gaugino condensation will
manifest itself in the constant $a$ in the exponent of $W_{D3I}$.
Note that the prefactor $C(z_i)$ is in general a complex structure
moduli dependent function. In order to be fully general, and also
in analogy to the heterotic case, we have included into $W$ a
third term, $W_S$, being an exponential in the type IIB dilaton
$\tau = iS$, and being again equipped with a complex structure
moduli dependent function $C'(z_i)$. This term could be motivated
by the action of the D(-1)-brane instanton [\ref{Green}]. (One
might also consider, here and in the heterotic theory, the
inclusion of NS 5-brane instantons wrapping $X$.)

{}From the K\"ahler potential (assumed to have the standard tree
level form)
\beqa
K=K(z_i,\bar z_i)-3 \log (T + \bar{T})-\log(S+\bar{S})
\eeqa
one finds that demanding unbroken supersymmetry in the complex
structure moduli $z_i$ sector yields the conditions
\beqa
\label{zcond}
D_i W = \int G \wedge \chi_i +\del_i C\, e^{-aT} + K_i \, W_{D3I}
\;\;\;\; \Big[ + \del_i C'\, e^{-bS} + K_i \, C'\, e^{-bS} \Big] =
0
\eeqa
(here $D_i\Omega=\chi_i$, with the $\chi_i$ being a cohomology
basis for (2,1)-forms). It follows that, with $W_{D3I}+W_\tau$
included, one needs $G^{1,2}\neq 0$ for unbroken
supersymmetry.\footnote{\label{overconstrain}Note that it seems
not possible in general to set all $G^{1,2}$-fluxes to zero, which
sets up a system of $n:=h^{2,1}$ equations for $n$ moduli fields,
and to simultaneously set up a cancellation among the other terms
in (\ref{zcond}) since this requires to solve another set of $n$
independent equations for the $z_i$.}

Similarly unbroken supersymmetry in the K\"ahler modulus sector
yields
\beqa
\label{Tcond}
D_T W = -aC\, e^{-aT}-\frac{3}{2t}W =0
\eeqa
Therefore one needs $G^{0,3}\neq 0$ for unbroken supersymmetry.
This Hodge component was implicitly already present in
[\ref{KKLT}] when a non-zero value $W_0$ of $W_G$ was discussed.
It goes beyond the supersymmetric flux components $G^{2,1}$ of
[\ref{GKP}] when discussing just $W=W_G$ but is still ISD. But as
we describe here the other two IASD components $G^{1,2}$ and
$G^{3,0}$ will be present as well.

Finally the dilaton sector shows the need of a $G^{3,0}\neq 0$ for
unbroken supersymmetry
\beqa
\label{Scond}
D_S W = -\frac{1}{2s}\int \overline{G}\wedge \Om -\frac{1}{2s}
W_{D3I} \;\;\;\; \Big[ -bC'\, e^{-bS}-\frac{1}{2s}C'\, e^{-bS}
\Big] =0
\eeqa

In conclusion, whereas supersymmetric vacua from a pure flux
superpotential are obtained from ISD 3-fluxes $G^{2,1}$, the $T$-
(and possibly $S$-) dependent corrections to $W$ imply that
supersymmetric groundstates correspond to more general ISD plus
IASD fluxes\footnote{Neglecting $W_S$ one gets $b^{1,2}_i = -K_i
\bar{\al} -\del_i C e^{-aT}/\frac{vol}{i}$ (adopting partially the
later notation (\ref{H types}) here for the {\em complex} $G$), or
$b^{1,2}_i=-K_i \bar{\al}$ (neglecting $\del_i C$) relating the
IASD components $G^{1,2}$ and $G^{3,0}$.} where all possible Hodge
types are turned on. (Specific type IIB orientifolds of this type
will be constructed in [\ref{Schulgin}].) The supersymmetric
minima obtained in this way are generically anti-de Sitter since
the groundstate has the property $W\neq 0$.

One may ask whether this result is also important for the
statistical counting of supersymmetric flux vacua [\ref{D}]. So
far in the literature the count for supersymmetric flux vacua was
done under the assumption that the 3-flux is ISD. Including the
corrections of the type discussed above will (as long as these are
suitably small) amount to tiny shifts of the critical points.

{\em Remarks on the use of the decoupling procedure}

Let us discuss these results in the light of the two-step
procedure applied in [\ref{KKLT}] for obtaining supersymmetric
groundstates from the superpotential (\ref{typeiiBsupo}). In the
first step groundstates of the pure 3-form superpotential $W_G$
were found for which the 3-form flux is purely ISD. More
precisely, as described in [\ref{GKP}], the conditions $D_i W_G =
0$ are solved by considering only fluxes with $G^{1,2}=0$ (the
Hodge-components $G^{0,3}$ and $G^{3,0}$ are similarly set to zero
if the conditions $W=0$ and $D_{\tau}W=0$ are imposed). This in
general fixes all complex structure moduli.

In the second steps one plugs in the values for the fixed complex
structure moduli into $W_G$ (and also into $C(z_i)$) and then
solves the supersymmetry conditions for the remaining K\"ahler
modulus $T$. As now a non-zero value $W_0$ of $W_G$ is employed
one has gone already beyond a pure $G^{2,1}$ and has turned on a
$G^{0,3}$, as already done in [\ref{KKLT}].

The work of [\ref{KKLT}] is based on this set-up and then
crucially assumes that the $z_i$ are much heavier than $T$ such
that they decouple and one merely remains with the stabilization
problem for $T$. However without the assumption of the $z_i$ being
integrated out, i.e. without assuming the KKLT two-step
stabilization procedure, the $D_iW=0$ condition picks up a further
contribution such as $K_i W_{D3I}$, which actually enforces a
non-trivial $G^{1,2}$ component.

So the above analysis shows that this two step procedure is only
of limited justification (similar conclusion were drawn in
[\ref{Nilles}]). The generic situation is more appropriately
captured by solving all supersymmetry conditions at the same time.
Then supersymmetric 3-form flux is not only ISD, but all Hodge
types appear.

The advocated two step procedure of [\ref{KKLT}] is only justified
if the $T$- (or $S$-) dependent corrections are indeed suitably
small, such that the flux is almost of Hodge type $(2,1)$, and the
complex structure moduli are fixed to values which are very close
to those of a pure 3-form flux superpotential. Here it will be
helpful to have many 3-cycles making possible to have an
exponentially small $W_{G}$.

Then let us call $W_0$ the value of $W_G$ for these moduli. This
is now only a constant, generically non-vanishing, which just can
occur in $K_S W$- or $K_T W$-parts of covariant derivatives of the
full $W$. So, starting from $W=W_0 + Ce^{-aT}$  with $a$ real and
$K^{(T)}=-3\log(T+\bar{T})$, one finds after the cancellation of
the $-3|W|^2$ part for the potential (with $T=t+i\si_T$)
\beqa
V=\frac{1}{8t^3}\Bigg(
\frac{1}{3}a\Big(a+\frac{3}{t}\Big)\Big(2t|C|e^{-at}\Big)^2 + 2t a
e^{-at} 2 \mbox{Re}(C \bar{W_0} e^{-aT}) \Bigg)
\eeqa
When adopting the often imposed condition $C, W_0 \in {\bf R}$ the
second term in the brackets becomes $aW_02tCe^{-at}2\cos a\si_T$.
This gives finally in the sector without axion-component
\beqa
\label{V reduced}
V|_{\si_T=0}=\frac{aCe^{-at}}{2t^2}\bigg(
\Big(\frac{at}{3}+1\Big) C e^{-at} + W_0 \bigg)
\eeqa

When employing this `integrating out' procedure\footnote{All of
this concerns just the part of the argument before a potential
`up-lift' (by $\overline{D3}$-branes, say).} a number of points
should be addressed:
\begin{itemize}
\item In principle one has to make sure that the shift in $z$,
caused by fixing it just by $W=W_G$ instead of using the full
$W=W_G + W_{D3I}$, is appropriately small (for remarks on this
cf.~[\ref{Dine}]). \item It must be checked that the stabilized
$z_i$ are more heavy than the stabilized $T$ modulus (in principle
this should be checked for the moduli values stabilized from the
full superpotential, cf.~the first point, not just for (\ref{V
reduced})). \item The stability of the stationary point has to be
checked; this is a non-trivial point and not always a minimum is
found [\ref{Nilles}].
\end{itemize}

\section{Heterotic Moduli Stabilization with 3-Form Fluxes
and Non-Perturbative Corrections}

\resetcounter

We will first recall the argument presented in [\ref{BdA}] for a
necessary emergence of the $H^{2,1}$ component and connect then to
the stabilization of the dilaton also discussed in [\ref{GKLM}].

\subsection {\label{3 flux het}3-Form Fluxes in the Heterotic String}

An important mechanism to break supersymmetry while still
maintaining a vanishing cosmological constant at tree level stems
from a complete square in the heterotic string Lagrangian
suggesting a cancellation between the gaugino condensate and an
$H$-flux [\ref{DRSW}]
\beqa
\int d^{10}x \sqrt{-g} \, \Big( H_{mnp} - \al' \tr \, \bar{\chi}
\Gamma_{mnp}\chi \Big)^2 \; .
\eeqa
The gauginos condense non-perturbatively such that $\tr \,
\bar{\chi} \Gamma_{mnp}\chi $ acquires a vacuum expectation value
(vev)
\beqa
\langle \, \tr \bar{\chi} \Gamma_{mnp}\chi \, \rangle = \La^3 \,
\Om_{mnp} + c.c. \; ,
\eeqa
where $\Om_{mnp}$ is the holomorphic $(3,0)$ form of the internal
Calabi-Yau manifold and $\La^3 = \langle \, \tr \bar{\la}_D
\frac{1}{2}(1-\gamma_5)\la_D \, \rangle = \langle \tr \la \la
\rangle$ the vev of the four-dimensional condensate. One gets
Minkowski vacua of vanishing tree-level potential for the flux
choice $H=H^{3,0}+c.c.$ with
\beqa
H = \al' \, \La^3 \, \Om + c.c.
\eeqa

The other Hodge component $H^{2,1}+c.c.$ can be turned on
supersymmetrically only if the underlying compactification
geometry possesses the balancing property of being non-K\"ahler
[\ref{S}], [\ref{CCdAL1}], [\ref{B1}] (cf.~also
[\ref{Gauntlett1}], [\ref{Gauntlett2}]) which allows for a
nontrivial $\del J \ne 0$ and thus
\beqa
H^{2,1}+c.c. = \frac{i}{2}\del J + c.c.
\eeqa
This balancing condition can also be obtained [\ref{CCdAL2}],
[\ref{B2}] from a superpotential of the form $W = \int_X
(H+\frac{i}{2} \, dJ) \wedge \Om$. In this framework the gaugino
condensate can be included via its effective superpotential as
well [\ref{CCdAL3}] (cf.~also [\ref{CCdAL4}]). Unfortunately, the
moduli space of non-K\"ahler spaces is still not appropriately
understood. For the issue of the stabilization of the complex
structure moduli $z_i$, K\"ahler modulus $T$ and dilaton $S$, it
would therefore be favorable to have separate control over the
$H^{2,1}$ and $H^{3,0}$ sectors. One starts from an effective
superpotential description of the $H$-flux and the gaugino
condensate in four-dimensional moduli fields, but it is not quite
clear what in the non-K\"ahler situation the `complex structure
moduli' and `K\"ahler moduli' really are. It is one of the goals
of the present work to show under which conditions we can
consistently set $H^{2,1}$ supersymmetrically to zero and still
stabilize all $z_i,T,S$ moduli in a reliable way, while keeping a
non-trivial $H^{3,0}$.

One quickly faces a problem since supersymmetry seems to require a
non-trivial $H^{2,1}$ once a non-vanishing $H^{3,0}$ is induced
through gaugino condensation. The situation  is the following:
when one is arguing via the potential obtained from dimensional
reduction, one finds
\beqa
\langle \tr \la\la \rangle \neq 0 \;\; \Lra \;\; H^{3,0}\neq 0
\qquad\quad \mbox{while} \qquad\quad H^{2,1}=0 = dJ
\eeqa
However in [\ref{BdA}] it was shown, working to lowest order in
$\al'$, that, in marked contrast to this result, {\em in an
effective four-dimensional supergravity approach}, incorporating
$H$-flux and gaugino condensation via the combined superpotential
\beqa
W=W_H[z_i] + W_{GC}[S]
\eeqa
all $z_i$ moduli and the dilaton $S$ could be fixed, but it was
not possible to turn on exclusively $H^{3,0}$ flux in a
supersymmetric way while having $H^{2,1}=0$
\beqa
\langle \tr\la\la \rangle \neq 0 \;\; \Lra \;\; H^{3,0}\neq 0
\qquad\quad \Lra \qquad\quad H^{2,1}\neq 0 \neq dJ \; .
\eeqa
The reason for this failure is recalled in subsect. \ref{het
sugra} where we also connect to the work of [\ref{GKLM}], pointing
to inclusion of the periods besides the flux. Let us note that the
vacuum found for the combined superpotential $W$ is
non-supersymmetric because one still has to take into account the
$T$ sector with $D_T W = K_T W \neq 0$. Since $\del_T W =0$ one
finds that $T$ remains undetermined. In this no-scale model the
vacuum energy is zero.

\subsection{The Potential From Dimensional Reduction}

Starting with a ten-dimensional metric (we follow conventions of
[\ref{GKLM}] which normalizes the Calabi-Yau metric $g^{CY}_{mn}$
to have volume $4\al'^3$)
\beqa
\label{metric}
ds_{10}^2 = e^{-6\si} ds_4^2 + e^{2\si} g^{CY}_{mn} dy^m dy^n \; ,
\eeqa
the dimensional reduction of the ten-dimensional action leads to
the complete square in the potential
\beqa
\label{compl squ}
V \sim \int_X d^6y \sqrt{-g^{CY}} \Big( H -
\frac{e^{12\si}}{16}l_{str}^2\T \Big)^2 \; .
\eeqa
Here one has the decompositions ($H$ being closed and harmonic for
the standard embedding of unbroken hidden $E_8$ with $C_G=30$, and
in general up to $\alpha'$ corrections)
\begin{alignat}{3}
\label{GC types}
\T &= 2 U\Om + c.c. \\
\label{H types}
H &= \al \Om + b^i \chi_i + c.c. \; , \quad i=1, \hdots,
h^{2,1}(X) \; ,
\end{alignat}
where the gaugino condensate is described by the effective field
\beqa
U = \langle \tr \la \la \rangle  = 16 \pi^2 m_{KK}^3e^{-\frac{2\pi
f_{hid}}{C_G}}= e^{-12\si} \mu^3 e^{-\frac{2\pi f_{hid}}{C_G}}
\eeqa
with $\Re f_{hid}=4\pi / g^2_{hid}$, the Kaluza-Klein scale given
by $m^3_{KK}=e^{-12\si}\frac{c}{2}m_{str}^3$, $c$ being an ${\cal
O}(1)$ numerical constant [\ref{GKLM}]) and
\beqa
\mu^3 = 8 \pi^2 \, c \, m_{str}^3 \; .
\eeqa
Furthermore one can define suitably contracted expressions
\beqa
b_i = G_{i\bar{k}}\bar{b}^{\bar{k}}\; , \quad b_{\bar{\jmath}}
= G_{l\bar{\jmath}}b^l
\eeqa
with the complex structure moduli space metric
\beqa
G_{i\bar{\jmath}} = - \frac{\int \chi_i \wedge
\bar{\chi}_{\bar{\jmath}}}{vol / i} \; , \quad vol / i = \int_X
\Om \wedge \bar{\Om} \; .
\eeqa

To relate expressions involving fluxes to periods over specific
cycles we choose a symplectic basis $(A^p, B_q)$ of $H_3(X,
{\mathbf Z})$ for which $A^p \cap A^q = B_p \cap B_q = 0$ and $A^p
\cap B_q = \delta^p_q$ with $p,q= 0 , \hdots , h^{2,1}(X)$. One
then defines the periods
\beqa
X^p= \int_{A^p}\Om \; , \quad F_q = \int_{B_q}\Om
\eeqa
and has the 3-cycle $C(H) = e_p A^p - m^q B_q$ dual to $H$. This
means the integrally quantized [\ref{Rohm Witten}] fluxes are
given by (our normalization differs by a factor of $\pi$ from
[\ref{Rohm Witten}]
\beqa
\label{dB fluxes}
\frac{1}{2\pi^2 l_{str}^2}\int_{A^p}H = m^p \;\;\; , \;\;\;
\frac{1}{2\pi^2 l_{str}^2}\int_{B_q}H = e_q
\eeqa
so that one finds
\beqa
\label{h full}
h = \frac{1}{4\pi^2 l_{str}^5}\int_X H\wedge \Om &=& -\,
\frac{1}{2 l_{str}^3} \;
\left( e_p X^p - m^q F_q \right)\\
&=&\frac{1}{4\pi^2 l_{str}^5} \; \bar{\al}\, \frac{vol}{i}
\eeqa
which represents roughly the number of flux quanta [\ref{GKLM}].

\subsection{Balancing Flux and Gaugino Condensation
}

Let us first assume that $H$ is supported on a 3-cycle $C_k$ with
integer flux $n_k$
\beqa
\frac{1}{2\pi^2 l_{str}^2}\int_{C_k} H_B &=& n_k
\eeqa
and period
\beqa
\Pi_k = \int_{C_k} \Om \; .
\eeqa
The gaugino condensate then contributes to the potential obtained
via dimensional reduction through (where $S=s+i\si_S=f_{hid}$)
\beqa
\int_{C_k} \frac{e^{12\si}}{16}\, l_{str}^2\, (2 \langle \tr \la
\la \rangle \Om + c.c.) = \frac{c \pi^2}{l_{str}}\, e^{-\frac{2\pi
s}{C_G}}\,
\Big( e^{-\frac{2\pi i \si_S}{C_G}}\Pi_k + c.c. \Big) \; .
\eeqa
Hence minimizing the complete square of the potential gives the
following balancing equation between flux and gaugino condensate
\beqa
\label{compl squ balanc}
\begin{array}{|c|}
\hline n_k = \frac{c}{2}\; e^{-\frac{2\pi s}{C_G}}\;
\left(
e^{-\frac{2\pi i \si_S}{C_G}}\Pi_k + c.c.
\right) \, l_{str}^{-3}
\\
\hline
\end{array}
\eeqa
which has the usual problem of balancing an exponentially small
right-hand side (at weak coupling) with an integer (here
$(e^{-\frac{2\pi i \si_S}{C_G}} \Pi_k + c.c.) l_{str}^{-3} \approx
1$ was assumed in [\ref{GKLM}]).

One possible way out is to use a possible non-integrality of $n_k$
arising from the Yang-Mills Chern-Simons term [\ref{GKLM}]
(including $\al'$ corrections; $H$ becomes non-closed). The
Lorentz Chern-Simons term was ignored in [\ref{GKLM}] but still
the full $H=dB + CS_L - CS_{YM}$ has to be solved (potentially
taking for $A_{YM}$ just the spin-connection (`spin in the gauge')
but deformed by a flat $A$ on a sLag-cycle). Alternatively one may
discuss, as we will do next in the effective supergravity
approach, the period and its potential ability to compensate the
exponential suppression, cf.~remarks after (\ref{h general}) and
before (\ref{fluxes periods}).\footnote{One might also use the
strongly coupled heterotic string, with the hidden boundary
stabilised near the singularity [\ref{CK2}], [\ref{BCK}],
ameliorating the balancing $n\sim e^{-(S-\gamma T)/C_G}$ by having
$S - \gamma T \approx 0$.}


\subsection{\label{het sugra}The Effective Supergravity Approach}

Let us now come to the effective four-dimensional description for
which we define from the metric (\ref{metric}) and the
ten-dimensional dilaton $\phi$ the real moduli
\beqa
s = e^{-(\phi/2 - 6\si)} \; , \quad t = e^{\phi/2 + 2\si} \; .
\eeqa
Together with the corresponding axions they build the complex
scalars
\beqa
S=s+i\si_S \;\;\;\; , \;\;\;\; T=t+i\si_T
\eeqa
and satisfy $e^{12\si} = s^{3/2}t^{3/2}$. The complex structure
moduli $z_i$ are defined via the periods
\beqa
z_i = \frac{X^i}{X^0} \; , \quad i=1,\hdots,h^{2,1}(X) \; .
\eeqa

The effective description describes the flux and gaugino
condensate effects through the superpotentials (actually a further
factor $e^{-1}$ occurs in the normalization of $W_{GC}$ in
[\ref{BdA}]; this will cause no difference in our argument)
\beqa
\label{supo}
W&=&W_H + W_{GC}\\
W_H &=& \frac{4}{\al'^4}\int H\wedge \Om
= \frac{4}{\al'^4} \, \bar{\al}\frac{vol}{i}
=\mu^3 \frac{2}{c}h \\
W_{GC}&=& -C_G \mu^3 e^{-\frac{2\pi f_{hid}}{C_G}} = - C_G \mu^3
e^{-\frac{2\pi S}{C_G}} \; .
\eeqa
Here $\Re f_{hid} = 4\pi / g^2_{hid} = s$ is the classical result
for the gauge kinetic function in the weakly coupled heterotic
string. Furthermore one has for the K\"ahler potential
\beqa
K = - \log 2 s - 3 \log 2 t - \log vol \; .
\eeqa

When searching for supersymmetric vacua one has to consider the
minimization of the corresponding scalar potential $V_{Sugra}$. Up
to a multiplicative K\"ahler factor, which is of no concern to us
here, the effective supergravity potential is ($D_i \equiv
D_{z_i}$)
\beqa
\label{sugra potential}
V_{Sugra} \sim K^{S\bar{S}}D_SW D_{\bar{S}}\bar{W} +
G^{i\bar{\jmath}} D_iWD_{\bar{\jmath}}\bar{W}
\eeqa
It will be interesting to note that the effective potential,
derived by dimensional reduction of the action (\ref{compl squ}),
\beqa
V \sim \Big|\al + \frac{l_{str}^2}{8C_G}
W_{GC}\Big|^2+G_{i\bar{\jmath}} b^i b^{\bar{\jmath}} \; .
\label{10DPot}
\eeqa
differs from the four-dimensional supergravity result
(cf.~[\ref{BdA}]).

Vacua with unbroken supersymmetry follow from setting $D_S W =
D_TW = D_iW = 0$. However, since our superpotential is $T$
independent, the model in the present case is of no-scale type,
which is the reason why the negative $-3|W|^2$ term cancelled.
This means that supersymmetry will be broken with non-negative
vacuum energy due to a non-vanishing F-term $D_T W = K_T W =
-\frac{3}{2t} W \neq 0$. As, without having included an explicit
$T$-dependence, the condition $D_T W=0$ can not be solved by a
finite $T$ we will therefore only demand that $D_S W = D_i W
=0$. The covariant derivatives are given explicitly as follows
\beqa
D_SW &=& -\frac{2\pi}{C_G}W_{GC}+K_S W_{GC} + K_S W_H \nonumber \\
        &=& K_S \Bigg[ \Big(\frac{4\pi}{C_G}s + 1 \Big) W_{GC} + W_H \Bigg]
\nonumber \\
D_iW &=& D_i W_H + K_i W_{GC}
\eeqa
giving finally, by using that $D_i \Om = \chi_i$,
\beqa
\begin{array}{|ccc|}
\hline D_SW &=&K_S \bigg( \left(\frac{4\pi}{C_G}s + 1\right)
W_{GC}
+ \frac{4}{\al'^4}\bar{\al} \frac{vol}{i} \bigg) \\
D_iW &=&\frac{4}{\al'^4} b_i  \frac{vol}{i} +  K_i W_{GC}
\\
\hline
\end{array}
\eeqa

Let us first focus on the dilaton equation $D_S W = 0$. It gives
the balancing between the flux-period product and the gaugino
condensate
\beqa
\label{DSW=0}
-\, \frac{8\pi^2}{\al'^3}(e_p X^p - m^q F_q) = \frac{4}{\al'^4}\,
\bar{\al} \, \frac{vol}{i}  &=&
-\Big(1+\frac{4\pi}{C_G}s\Big)\, W_{GC}\nonumber \\
&=&\mu^3 (C_G+4\pi s)\, e^{-\frac{2\pi S}{C_G}} \; .
\eeqa
Hence, as compared to (\ref{compl squ balanc}), the effective
supergravity approach leads to a (well-known) additional factor in
the flux condensate balancing equation
\beqa
\label{h general}
\begin{array}{|c|}
\hline h = -\, \frac{1}{2 l_{str}^3} \left( e_p X^p - m^q F_q
\right)
=\frac{c}{2}\, (C_G + 4\pi s)\; e^{-\frac{2\pi i \si_S}{C_G}}
e^{-\frac{2\pi s}{C_G}}
\\
\hline
\end{array}
\eeqa

{\em Remark:} To recover (\ref{compl squ balanc}) one may evaluate
$\int_X H\wedge \Om$ under the assumption that, schematically, $H$
is supported just on $C_k$ (of associated dual cycle $D_k$ in a
symplectic basis). Compared with (\ref{compl squ balanc}) the
inverse of the dual period appears because one has
\beqa
\label{h schematic}
h = \frac{1}{4\pi^2 l_{str}^5}\int_X H\wedge \Om = n_k \cdot
\frac{1}{2l_{str}^3}\int_{D_k} \Om \; ,
\eeqa
i.e. one gets in total
\beqa
\label{sugra balanc} n_k = \frac{c}{2}\, (C_G + 4\pi s)\;
e^{-\frac{2\pi s}{C_G}}\;
\Big(\frac{1}{2l_{str}^3} \int_{D_k} \Om \Big)^{-1} \; .
\eeqa
Using, however, the full expansion (\ref{h full}) instead of
(\ref{h schematic}) one gets a matching from (\ref{h general}) (up
to the $C_G + 4\pi s$ factor). To address the fact that in the
weakly coupled regime the right-hand side of (\ref{h general}) is
much smaller than one, one needs to check the size of the periods
appearing on the left-hand side. These are linked to the size of
the complex structure moduli such that the questions of
stabilization of the complex structure moduli and $s$ have to be
treated together.

\subsection{Emergence of the $H^{2,1}$ Component}

Let us next come to the complex structure equation $D_i W = 0$,
from which one gets
\beqa
\frac{4}{\al'^4}\, b_i \, \frac{vol}{i} = - K_i W_{GC} \; .
\eeqa
Together with (\ref{DSW=0}) this implies the following fixing of
the $z_i$
\beqa
\label{b_i fixing}
\begin{array}{|c|}
\hline b_i\Big(1+\frac{4\pi}{C_G}s\Big) = \bar{\al}K_i
\\
\hline
\end{array}
\eeqa
Hence one arrives at the conclusion [\ref{BdA}] that $b_i \neq 0$
and therefore $H^{2,1}\neq 0$, as otherwise $\alpha$ and therefore
$H^{3,0}$ would have to vanish or $s \rightarrow \infty$.

Note that in the given description of complex structure moduli
stabilization it is assumed that (as is has yet to be determined
what the consistent complex structure of the underlying Calabi-Yau
space $X$ actually is) the $H$ is given as an input as a real
three-form; then one has to rotate within the possible complex
structures of $X$ until one finds $H^{2,1}=0$. From this one
obtains a fixing of the $z_i$ by using one of the following two
reasonings: either as demanding $b_i=0$ poses $n=h^{2,1}$
conditions on the $z_i$ (this parallels the similar type IIB
argument) or exploiting [\ref{Rohm Witten}] the ensuing
proportionality between $H$ and $\Om + c.c.$, and the integrality
of $H=dB$ (if no $CS$-terms become manifest). Having fixed already
the $z_i$ there is no room to satisfy the further conditions $K_i
\, \bar{\al}/(1+\frac{4\pi}{C_G}s)=0$ on the $K_i(z_j)$, and so on
the $z_j$, what leads to the contradiction
(cf.~footnote~\ref{overconstrain} for the IIB case).

So in this framework one finds the interesting feature that it is
not possible to turn on in a supersymmetric way (this concerns the
$z_i$ and $S$ sectors) an $H^{3,0}$ flux while keeping
$H^{2,1}=0$. This is in marked contrast to the conclusion which
would be reached arguing just from the potential coming from
dimensional reduction, the difference coming from the $K_i W_{GC}$
term in the second term in (\ref{sugra potential}) (the different
$(\frac{4\pi}{C_G}s + 1)$ factor coming from the covariant
derivative $D_S$ is for this question not essential).

\subsection{Implementing a KKLT-Like Two-Step Moduli Stabilization Procedure}


In analogy to the type IIB case one can now contemplate the
possibility of using the two-step procedure
integrate out the $z_i$ in the problem first, i.e. from just $D_i
W_H=\int H \wedge \chi_i=\bar{b_i}=0$ such that $H$ is
$(3,0)+c.c.$ and the $z_i$ are fixed, following [\ref{Rohm
Witten}], from the ensuing integrality of (a multiple of) $\Om +
c.c.$. So by implementing the KKLT-like two-step procedure of
moduli stabilization, assuming the $z_i$ are heavy
enough\footnote{If the mass of the $z_i$ lies at or above the
threshold given by the inverse size of the Calabi-Yau space, then
their stabilization has to be discussed within the ten-dimensional
theory. In this case one also sees from (\ref{10DPot}) that
$b^i=0$ fixes the $z_i$.} to be integrated out first, the
emergence of $H^{2,1}+c.c.$ could be avoided.

{\em The problem of stabilization at weak coupling and the
heterotic discretuum}

Finally let us comment on the question of stabilisation between
the flux/period-product and the exponentially suppressed gaugino
condensate expression in (\ref{h general}). In type IIB one needs
analogously a hierarchically small value of $W_0=W_G$ to have $T$
fixed at large volume; similarly here a not small but $\O (1)$
flux value $n_k$ in (\ref{compl squ balanc}) prohibits a balancing
with a large $S$, i.e. being at weak coupling. One possibility is
to invoke the fractional flux argument [\ref{GKLM}] where $H$ is
no longer closed as one will not be in the case of the standard
embedding where the Chern-Simons terms would cancel. For other
proposals cf.~for example [\ref{NN}], [\ref{X}]. In the spirit of
the investigations about type IIB string theory related to the
discretuum one would argue for a sufficiently small value of $W_H$
(exponentially suppressed  for a weak coupling solution for $s$)
just from a heterotic discretuum [\ref{BdA}] for large $h^{2,1}$
like in type IIB. But note that heterotically there are only half
as many fluxes per modulus, making tunability nearly
impossible.\footnote{We thank the referee for bringing this point
up.} We however expect that the completion of $H$ by $dJ$ makes
for a fuller analogy with type IIB even in this respect.

Note that a reasoning for a near conifold vacuum as in type IIB
[\ref{GKP}], [\ref{D}], to make even at least one period in $W_H$
sufficiently small, meets the following obstacle caused by the
reality of $H$: with the vanishing cycle $A= {\bf S^3}$ and dual
cycle $B$
\beqa
\label{fluxes periods}
\int_A H   = M \;\;\; &,& \;\;\; \int_B H = K \\
\int_A \Om \; = \; z \;\;\; &,& \;\;\; \int_B \Om = \G(z)
=\frac{1}{2\pi i}z \ln z + \mbox{holo.}
\eeqa
(with integers $M$ and $K$) one has for the flux superpotential
\beqa
W_H = \int H \wedge \Om = M \G - K z
\eeqa
Then one finds for $K \gg 1$ from the condition for the complex
structure modulus $z$
\beqa
\label{z equ}
0=D_z W_H = M \G'-K + K_z W_H \approx \frac{M}{2\pi i}\ln z - K +
{\cal O}(1)
\eeqa
(imposing the $D_z W = 0$ condition just for the $W_H$ sector
assumes the two-step procedure).

Therefore $z \approx e^{2\pi i K/M}$ rather than being
exponentially small. (For a supersymmetric BPS cycle $A$ (in type
IIB) the sLag condition shows $\int_A \Om$ is just a real volume.)
So the reality of $H$ prohibits a relation $\int_B H = iK$, needed
to stabilize a near-conifold vacuum, which was possible in type
IIB where $\int_B G = iK/g_s$ (for a purely imaginary dilaton).

In [\ref{CCdAL2}], [\ref{CCdAL3}], [\ref{B2}] a complex version of
$H$ was used
\beqa
{\cal H}=H + \frac{i}{2}dJ
\eeqa
and led to satisfying results for the corresponding superpotential
when compared with the potential coming from dimensional
reduction. By contrast in the present paper we put $dJ=0$ and use
the effective four-dimensional supergravity approach where the
appropriate moduli space in the case $dJ\neq 0$ is not yet well
understood. Using the imaginary component of ${\cal H}$ would one
bring precisely back to that problem of non-K\"ahlerness.
Nevertheless by invoking this possibility one would make an even
closer analogy to the type IIB case as one gets a size-dependent
imaginary part of ${\cal H}$ and thus the analogue (under the
substitution $\tau \ra T_{het}$ and $T_B\ra S_{het}$ which relates
heterotic case and type IIB, cf.~[\ref{Nilles}]) to the type IIB
three-form
\beqa
G = F - \tau H
\eeqa

\section{\label{WSI sect}Inclusion of a
Non-Perturbative Size-Fixing Superpotential}

\resetcounter

Above the superpotential did not yet depend on the $T$ modulus,
i.e. it was of no-scale type and $T$ remained unfixed. Although a
volume-modulus stabilizing effect of a non-trivial $H^{2,1}$
together with $dJ$ is known [\ref{S}], [\ref{CCdAL2}], [\ref{B1}]
it is the latter component which we wish to avoid here for the
mentioned reason of having a clear-cut notion of K\"ahler- and
complex structure moduli. In the similar framework of the strongly
coupled heterotic string the inclusion of a non-perturbative
size-fixing superpotential led to a stabilization of the $T$
modulus [\ref{CK2}], [\ref{BCK}]. There the $T$ modulus was
stabilized by balancing the effects of a non-perturbative
size-fixing superpotential (induced from open membrane instantons)
$e^{-T}$
with the contribution $e^{-f_{hid}/C_G}$ coming from the gaugino
condensate (a flux superpotential $\int_X H\wedge \Om$ can there
also be included). Similarly here we will discuss how the
inclusion of world-sheet instantons\footnote{Actually only the
product $Be^{-iT_I}$, and not these factors individually, is
strictly well-defined [\ref{W}].} $W_{WSI}= B \, e^{-T}$ changes
the above results (cf.~also [\ref{BO}]). For its effectivity we
assume non-standard embedding\footnote{\label{non-stand embed
definition} If one then wants, having included an effect
non-perturbative in $\al'$, to take into account perturbative
$\al'$ corrections so that $dH\neq 0$ one replaces (\ref{H types})
by $H=H^{3,0}+H^{2,1}+c.c.$ and {\em defines}\, $\bar{\al}\,
vol/i:= \int H \wedge \Om$,\, $b_i vol/i:= \int H \wedge
\bar{\chi_i}$ (after having chosen representatives) so that
demanding $H^{2,1}=0$ still implies $b_i=0$.} (otherwise
(\ref{Pfaff})-terms sum to zero). We will first discuss whether it
is possible to have $H^{2,1}=0$ without using the two-step
procedure mentioned above.

The complete square structure is now broken by $\del_T W \neq 0$:
the $K^{T\bar{T}}D_TWD_{\bar{T}}\bar{W}$ term of $V_{sugra}$ will
not just cancel the $-3|W|^2$ term. Nevertheless stationary points
of $\del V=0$ can, for supersymmetric vacua, still be found from
just the $DW=0$ conditions.

To demonstrate the unavoidability of the new Hodge type $H^{2,1}$
we ask whether one can find now susy vacua with $b_i=0$. For this
we start from the following superpotential (assuming $h^{1,1}=1$)
\beqa
W&=&W_H + W_{WSI} + W_{GC}\nonumber\\
    &=&\int H\wedge \Om + B e^{-bT} + C e^{-aS}
\eeqa
One gets (with the notation $W^{S,T} = W_{WSI} + W_{GC}$)
\beqa
\label{i cond}
D_i W &=& b_i \, \frac{vol}{i} + \del_i B \, e^{-bT} + K_i W^{S,T}\\
\label{S cond}
D_S W &=& -aCe^{-aS} - \frac{1}{S+\bar{S}}W
= - \frac{1}{2s} \bigg( 2as \cdot Ce^{-aS} + W_H + W^{S,T} \bigg)\\
\label{T cond}
D_T W &=& -bBe^{-bT} - \frac{3}{T+\bar{T}}W
= - \frac{3}{2t}
\bigg( 2\frac{bt}{3} \cdot Be^{-bT} + W_H + W^{S,T} \bigg)\
\eeqa

Let us at first in (\ref{i cond}) neglect the contribution $\del_i
B \neq 0$ (this will be remedied below). Then imposing in addition
the condition $b_i=0$ gives $W^{S,T}=0$, i.e.
\beqa
\label{WST = 0}
Ce^{-aS}=-Be^{-bT}
\eeqa

So here the complex structure moduli are fixed by $b_i=0$.
Independently of the explicit value $W_0=\bar{\al}\, vol/i$ of
$W_H$ (\ref{S cond}) and (\ref{T cond}) give already a relation
between $S$ and $T$
\beqa
\label{S T cond}
as \cdot Ce^{-aS}=\frac{bt}{3}\cdot Be^{-bT}
\eeqa
(\ref{WST = 0}) and (\ref{S T cond}) determine\footnote{The
earlier case is included formally as the degenerate case $B=0$ not
corresponding to a finite $s$.} $s$ and $t$ but outside the regime
of physical validity as the ensuing relation $as = -bt/3$ between
the $a,b,s,t >0$ shows; so one can not have $b_i=0$.

With $\del_i B$ included one has from the demand $b_i=0$ now
instead of $W^{S,T}=0$ that
\beqa
\label{implication}
b_i=0 \;\;\;\; \Longrightarrow W^{S,T}=-\frac{\del_i B}{K_i}
e^{-bT}
\eeqa
i.e. one finds instead of (\ref{WST = 0}) now
\beqa
\label{WST = 0 completed}
Ce^{-aS}=-\bigg( \frac{\del_iB}{B\, K_i}+ 1 \bigg) Be^{-bT}
\eeqa

Note that the interpretation would now be different than before.
In the previous section we considered the $H$-flux as a given real
form and a fixing of the $b_i$ as in (\ref{b_i fixing}) was
considered as fixing the complex structure moduli $z_i$. Now we
would have to regard $H$ as adjustable, impose the demand $b_i=0$
and get as conditions for the $z_i$ the relations (\ref{WST = 0
completed}) where one gets for the $z_i$ (still entangled there
with the $S$ and $T$ yet to be determined) the conditions (for all
$i=1, \dots , h^{2,1}$; the constant $k$ defined on the solution
pair $(S,T)$)
\beqa
\label{k parameter}
\frac{\del_iB}{BK_i}+1 =k:= - Ce^{-aS}/Be^{-bT}\, .
\eeqa

So essentially this condition on the $\del_i B/(BK_i)$ would
determine the $z_i$ (i.e. modulo the coupled determination of the
$S$ and $T$).\footnote{In principle one could have tried to use
this reasoning to solve (\ref{b_i fixing}) with the conditions
$K_i=0$ for the $z_j$ and a similar adjustment of $H$ as here; but
without $W_{WSI}$ one would not have fixed the $T$ modulus as it
does otherwise the $H^{2,1}$.} Now (\ref{S T cond}) gives, with
$a, b > 0$, the condition $C e^{-aS}/(B e^{-bT})\in {\bf R^{>0}}$,
i.e. $k\in {\bf R^{<0}}$; the point here is that previously,
without $\del_i B$, the parameter $k$ was just $1$ leading to a
contradiction to $k\in {\bf R^{<0}}$.
So one would now have consistently $b_i=0$ in the sense that for a
given $\al$ the $S,T$ and the $z_i$ are determined with having
also $b_i=0$. (\ref{S cond}) and (\ref{T cond}) show that, as
$W_H\neq 0 $, there exists a second solution of finite $s$ and $t$
besides the runaway solution $s, t \rightarrow \infty$.

So the inclusion of $W_{WSI}=Be^{-bT}$ would make it possible,
taking into account that $\del_{z_i}B\neq 0$, to solve
consistently with $H^{2,1}=0$ {\em if a $H^{3,0}\neq 0$ could be
suitably adjusted}.

\subsection{Remarks on the Problem of Adjusting the $H$-Field}

When including world-sheet instanton effects one has to work
outside the standard embedding as otherwise contributions coming
from the curves in the same cohomology class would sum up to zero.
So (with $\om_{YM}=\tr (AdA+\frac{2}{3}A^3)$ being the
Chern-Simons form)
\beqa
\label{Bianchi}
H=dB+CS \;\;\; \mbox{where} \;\;\; CS =
\frac{\al'}{4}(\om_{L}-\om_{YM})\;\; \Lra \;\; dH =
\frac{\al'}{4}\tr (R\wedge R - F\wedge F)
\eeqa
With $F\neq R$ now $H$ will also no longer be just $dB$ and
so\footnote{understood as a four-form; considering just cohomology
the integral of this closed four-form over a four-cycle gives the
number of fivebranes wrapping the dual two-cycle necessary for
anomaly cancellation} $dH\neq 0$ generically and $H$ can not be
decomposed as a sum (\ref{H types}) of closed forms. As relevant
for the $DW=0$ conditions were only the integrals $W_H = \int H
\wedge \Om$ and $\int H \wedge \bar{\chi_i}=0$ we just call, even
without a decomposition (\ref{H types}), the values of these
integrals $\bar{\al}$ and $b_i$ (cf.~footn. \ref{non-stand embed
definition}).

The change of interpretation mentioned after (\ref{WST = 0
completed}) deserves more discussion. Usually, when fixing the
$z_i$ by a flux, $H$ is given first as a real form on the
underlying real manifold and then the complex structure is fixed
as conditions emerge from $D_i W=0$. For example, if $H$ would be
built from working in the standard embedding $F=R$ such that
$H=dB$ and so $H$ is closed, a condition like $b_i=\int H \wedge
\bar{\chi_i}=0$ would amount to an equation $H=\al \Om + c.c.$
such that $\al \Om$ has to have integral periods what fixes the
$z_i$ [\ref{Rohm Witten}].

However above a different reasoning was tried: the $z_i$ were
fixed from the remaining (after putting $b_i=0$ in $D_iW=0$)
condition (\ref{WST = 0 completed}) and an $H$ of $b_i=0$ was
treated as if adjustable independently. So we have to ask whether
it is possible to turn on\footnote{if also $H^{3,0}=0$ the
equations (\ref{S cond}), (\ref{T cond}) have no finite solution
in $(s,t)$ as is easily seen} an $H=H^{3,0}+c.c.\neq 0$ {\em
without posing thereby additional conditions on the $z_i$ besides}
(\ref{WST = 0 completed}).

Actually, just when $H=H^{3,0} + c.c. = (dB)^{3,0} + (CS)^{3,0} +
c.c.$, one finds that $H$ is already closed\footnote{as the
mentioned Hodge-types for $CS$ give under $d$ the types $(3,1) +
c.c.$ and have therefore to vanish as $d CS = \tr (R\wedge R -
F\wedge F)$ has type $(2,2)$. Note further that $dH=0$ with $H\neq
0$ implies $\tr F\wedge F= \tr R\wedge R$ with $F\neq R$. To
potentially achieve this one may consider $F$ corresponding to
deformations $V$  of $TX$, cf.~the deformations $Q$ of $Q_0=TX
\oplus {\cal O}$ in [\ref{Witten 86}]. This uses the expansion in
$\al'/t$ when discussing corrections to the equations of motion
and the solvability of (\ref{Bianchi}) in cohomology (i.e.
assuming $c_2(V)=c_2(X)$, the absence of five-branes) is then
sufficient to solve (\ref{Bianchi}) for $H$ as form.}, i.e. $H=\al
\Om + c.c.$. But some $dB^{2,1}$ and $CS^{2,1}$ may cancel here.
Note that one can not split usually the contributions of $dB$ and
$CS$ as even the neutral field $B$ gauge transforms to achieve
invariance of $H$ under gauge transformations
\beqa
\de A = d \La + [ A, \La ] \longrightarrow \de \om_{YM}= d \, \tr
\, \La dA \;\;\; \Lra \;\de B = \frac{\al'}{4}\, \tr \, \La dA
\eeqa
One needs a non-trivial $A$ as $dB=\al_{dB}\Om + c.c.$ would have
to be integral and fix the $z_i$.

{\em On the use of CS-fluxes through special Lagrangian 3-cycles}

We discuss whether one can use advantageously fluxes through
special Lagrangian 3-cycles, cf.~[\ref{GKLM}]. For this we ignore
first the $dB$- and $\om_L$-part (possibly balancing the Bianchi
identity with additional five-branes) and view the remaining
$CS$-fluxes of Hodge-type $(3,0)+c.c.$ as fluxes through special
Lagrangian $3$-cycles $Q$ (cf. [\ref{GKLM}]).

A Lagrangian $3$-cycle $Q$ of $J|_Q=0$ is special Lagrangian
(sLag) if also $\Im\, \Om$ restricts to zero on it. On such
cycles\footnote{\label{slag}Examples are provided by the real
points of a Calabi-Yau manifold $X$ with possesses a real
structure (an antiholomorphic involution $\tau$ with $\tau J = -
J$ and $\tau \Im\, \Om = - \Im \, \Om$; the real points of $X$ are
the fixed points of $\tau$). We restrict the attention to {\em
rigid} special Lagrangian $3$-cycles $Q$ which are then known to
have $b_1(Q)=0$. Locally near such a cycle the Calabi-Yau geometry
of $X$ looks like $T^* Q$ and globally such a $Q$ could occur as
the base of a fibration $\pi: X \ra Q$ of $X$ by special
Lagrangian three-tori.} $\Om$ restricts to a multiple of the
volume form
\beqa
J|_Q=0 \;\;\;\; , \;\;\;\; \Im\, \Om|_Q = 0 \;\;\;\;\;\;\;\; \Lra
\;\;\;\; \Re \, \Om|_Q = \ga \, vol_Q
\eeqa

Now, how can one realize a closed $H=H^{3,0}+c.c.\neq 0$, i.e.
$H=\al \Om + c.c.$ (and not purely $dB$)? For this let $C_Q$ be a
real  three-form supported on a rigid sLag-cycle $Q$, i.e.
$C_Q=h_q\, vol_Q$ with $h_Q$ real. With $\Om|_Q = \ga \, vol_Q$
one finds, as $\frac{1}{\ga}\, \Om|_Q$ is a real form, indeed
\beqa
C_Q = h_q \, \frac{1}{\ga}\Om|_Q\;\;\;\; \;\;  \Lra \;\; \;\;\;\;
\int_Q C_Q \sim \int_Q H =\int_Q \al \Om + c.c.
\eeqa
($\pi$ as in footnote \ref{slag}) the latter from the Hodge-type
of the closed form $H$.

The sole use of a flat gauge connection $A$ to build $H$ would
still be insufficient as one has also for a pure
$H=-\frac{\al'}{4}\om_{YM}$ with $A$ flat a (fractional)
quantization like in (\ref{dB fluxes})
\beqa
\frac{1}{2\pi^2 \al'}\int_Q H = -\frac{1}{8\pi^2}\int_Q \om_{YM}
\in \frac{1}{p}{\bf Z}/ {\bf Z}
\eeqa
(($p\in {\bf Z}$); the integral being a topological invariant on
the moduli space of flat connections). $H\sim \om_{YM}$ would then
fix again the $z_i$ as did before the quantized $H=dB$.

So interpreting the Hodge-type condition $(3,0)+c.c.$ of the
sought-after $H$-flux as a condition of being such a $(\pi^*)C_Q$,
i.e. being supported on a sLag 3-cycle $Q$, one could be able to
turn on a non-flat $A$ on $Q$ without imposing thereby forbidden
additional conditions on the $z_i$ (which might have originated
from the Hodge-type restriction)
\beqa
dB + CS = h_q vol_Q
\eeqa
But turning on $\om_{YM}$ on $Q$ shows just how a Hodge-type
$(3,0)+c.c.$ could occur in general and makes not clear how to
satisfy the supersymmetry conditions $F=F^{1,1}$ with
$g^{i\bar{j}}F_{i\bar{j}}=0$ for $F\neq 0$ or how to obtain a
consistent package $H=CS=H^{3,0}+c.c.$ ($\om_L$ is ignored in
[\ref{GKLM}]). {\em Because of this difficulty we assume the KKLT
two-step procedure} where terms $K_i W^{S,T}$ and $\del_iB
e^{-bT}$ in $D_i W=0$ (and the first equation in (\ref{k
parameter})) do not arise.

\subsection{One-Loop Corrections}

Above it was assumed that the only $T$ dependence comes from
$W_{WSI}$, neglecting one-loop corrections [\ref{FKLZ}] to
$f_{hid}$ and effects from warp factors
$e^{12\si}=s^{3/2}t^{3/2}$. Both of these assumptions would have
to be modified when considering the strongly coupled heterotic
string.

When $H$ was decomposed in Hodge-types in (\ref{H types}) $\al'$
corrections were neglected so that $H$ could in particular assumed
to be closed; this starting point has to be modified if either
one-loop corrections to $f_{hid}$ are included in the weakly
coupled heterotic string framework or if the strongly coupled
heterotic string is discussed. In the latter case the $f_{hid}=
S+\beta T$ with $\beta = {\cal O}(1/100)$ of the weakly coupled
case is replaced by $S + \gamma T$ with $\gamma = {\cal O}(1)$ and
this `correction', which in the strongly coupled case is
inevitable, is directly related to the non-trivial part of the
Bianchi identity for $H_3$ resp. $G_4$, i.e. its non-closedness.
Note that a $T$ dependence in $f_{hid}$ which in the weakly
coupled heterotic string is included as a one-loop correction is
in the strongly coupled heterotic string included already by the
variation of the Calabi-Yau volume along the $x_{11}$-interval
[\ref{HetMWarp}].

We show now that the inclusion of $W_{WSI}$ will solve a problem
which arose in the analysis of [\ref{GKLM}]. First in a vacuum
without five-branes one has from the condition
$c_2(V_1)+c_2(V_2)=c_2(TX)$ for the one-loop corrections to the
gauge-kinetic functions
\beqa
f_{obs/hid}=S \mp \beta T
\eeqa
where $\beta=\frac{1}{8\pi^2}\int J \wedge
\big(c_2(V_2)-c_2(V_1)\big)$. The combined conditions $D_SW=0=D_TW$
in the case without $W_{WSI}=Be^{-bT}$ gave then in [\ref{GKLM}]
the condition
\beqa
\label{beta problem}
3s=\beta t
\eeqa
This caused a problem: besides the fact that one needs to have
$\beta > 0$ so that the observable sector is more strongly coupled
than the hidden one, one finds as a more serious consequence a not
well-understood strong coupling transition as one gets a negative
$\Re f_{obs} = -2s < 0$.

This problem is avoided in our approach as this time (\ref{beta
problem}) is replaced by\footnote{Note that $k$ is defined by the
second equation of (\ref{k parameter}) and the first equation of
(\ref{k parameter}) is absent in the two-step procedure employed
now.}
\beqa
\label{beta no problem}
\begin{array}{|c|}
\hline 3s=\left( \beta  - \frac{b}{ak}\right) t
\\
\hline
\end{array}
\eeqa
(where now actually $k:=-W_{GC}/W_{WSI}=-Ce^{-a(S+\beta
T)}/Be^{-bT}$). Here we found before, with $\beta=0$, from
(\ref{beta no problem}) that $k\in {\bf R^{<0}}$. Now we see that
one may solve both mentioned problems of [\ref{GKLM}]: we can
choose now the more standard choice $\beta < 0$,
and have from $\Re f_{obs} = -2s - \frac{b}{ak} t$ now that one is
not necessarily led to the strong coupling transition because $\Re
f_{obs}$ can now be positive for $k\in {\bf R^{<0}}$; more
precisely this happens for $2a\, |k|\, s < bt$ (demanding also the
stronger $\Re f_{hid} > 0$ gives $b/a|k|>4|\beta |$).

To corroborate these claims, let us start from the full
superpotential
\beqa
W &=& W_H + W_{WSI} + W_{GC}\\
&=& \int H\wedge\Om + B e^{-bT} + C e^{-a(S+\beta T)} \; ,
\eeqa
which includes the one-loop correction to the gaugino condensation
superpotential. Our goal is to solve
\beqa
\label{DW1}
D_S W &=& K_S \Big( 2as \, W_{GC} + W \Big) \; = \; 0\\
\label{DW2}
D_T W &=& K_T \Big(2\frac{bt}{3}\, W_{WSI} + 2 \frac{a\beta
t}{3}\, W_{GC} + W\Big) \; = \; 0  \; .
\eeqa
To this end, let us subtract the first from the second equation
with the result
\beqa
\label{Diff}
\frac{bt}{3}W_{WSI} = a\Big(s-\frac{\beta t}{3}\Big)W_{GC} \; ,
\eeqa
which is nothing but (\ref{beta no problem}). To stay within the
weakly coupled heterotic string regime, we have to impose $s\gg
|\beta t|$ and check its consistency later at the critical point.
Both prefactors are therefore positive. Moreover, to trust the
effective supergravity analysis, both $s$ and $t$ should be
considerably larger than one. We can then approximate both sides
of the equation just through the superpotentials alone (the
prefactors written as exponentials contribute only logarithmically
instead of linearly to the exponent and can hence be neglected at
sufficiently large $s$ and $t$). This leads\footnote{The balancing
of gaugino condensation with open membrane instantons is known
from the strongly coupled heterotic string to lead to a
stabilization of the orbifold-length (dilaton) [\ref{BCK}]. Indeed
for an unbroken hidden $E_8$ (as opposed to a hidden gauge group
of much smaller dual Coxeter number) the orbifold-length becomes
thus stabilized at rather small values close to weak coupling.
Here the open membrane instantons become heterotic world-sheet
instantons and we arrive at the balancing between $W_{WSI}$ and
$W_{GC}$.} to $W_{WSI}\simeq W_{GC}$, hence fixing $k\simeq -1$,
and gives us, by a similar reasoning as before suppressing the
prefactors in front of the exponentials, the relation
\beqa
\label{critical point}
\Big(\frac{b}{a}-\beta\Big)T \simeq S \; .
\eeqa

We can now adopt (\ref{DW1}) as the second equation determining
$S$ and $T$ besides (\ref{Diff}). Within the same approximation as
before it leads to $W_{GC}+W_H \simeq 0$ which is essentially
identical to (\ref{h general}) and can be solved with large $s$
either by having fractional flux [\ref{GKLM}] or by making use of
the heterotic discretuum [\ref{BdA}], probably allowing for
non-K\"ahlerness of the background, as discussed in section 3.

Two comments are now in order. First, for a hidden $E_8$ gauge
group we find $a=2\pi/C_G=2\pi/30$ whereas $b=1$. Thus, adopting a
value for $\beta$ of $\O(1/100)$, as appropriate for the weakly
coupled heterotic string, we obtain from (\ref{critical point})
that $5T \simeq S$ at the critical point. Hence, at the critical
point the one-loop correction to $f_{obs/hid}$ becomes $\beta T =
\O(S/500)$ which is indeed much smaller than the tree level result
$S$, showing that the critical point is consistently located in
the weakly coupled regime. Notice that this is not the case when
one neglects the world-sheet instanton superpotential as then the
critical point becomes characterized by (\ref{beta problem}) which
implies a one-loop ``correction'' $\beta t$ which is three times
as large as the tree level result $s$. The critical point in this
latter case is therefore situated outside the weakly coupled
string regime. Second and closely related to this first comment
about the smallness of the one-loop correction in the case with
world-sheet instantons, we do obtain a positive $\Re f_{obs}>0$ at
the critical point where
\beqa
\Re f_{obs} = s - \beta t =
\Bigg(\frac{\frac{b}{a}-2\beta}{\frac{b}{a}-\beta}\Bigg) s \; .
\eeqa
Since $\beta=\O(1/100)$ is much smaller than $b/a\simeq 5$ the
value is clearly positive. This is again in contrast to the case
without world-sheet instanton contribution for which $\Re
f_{obs}=-2s$ came out to be negative.

Hence the inclusion of $W_{WSI}=Be^{-bT}$ makes it also possible
(besides $T$ stabilization without $H^{2,1}$) to avoid the
occurrence of the uncontrolled transition found in [\ref{GKLM}]
while keeping the 1-loop correction small.

{\em Stability of the vacuum}

Finally we check stability of these vacua (for parameter values
where the axions $\si, \tau$ at the solution are zero) via the
criterion of [\ref{Nilles}]. With $W_{eff}^T[T] = W_H + Be^{-bT}$
from (\ref{W eff}) the stability parameter $\eta = t W_{eff}^{T\;
\prime\prime} / W_{eff}^{T\; \prime} $ becomes $\eta = -bt$ and
the stability criterion
\beqa
| \eta - 1 | = | bt + 1 | > 1
\eeqa
(in leading order in $1/(as)$) is with $b > 0$ automatically
fulfilled. Actually and more appropriately here, because of the
relation $\eta = 3ask$ following from (\ref{S T cond}) and (\ref{k
parameter}), the reasoning comparing orders in $1/(as)$ has to be
slightly reconsidered; the conclusion of stability, in the sector
$as \gg 1 $ relevant for us, remains unchanged.

The inclusion of the previously mentioned one-loop corrections
should not change this result as they have to be small and hence
cannot change the sign of $V''$.

\noindent {\em Remark 1: The possibility of flux-less solutions}

With the inclusion of the mentioned one-loop effect it becomes
possible to solve the $DW=0$ conditions consistently even with
$b_i=0$, without having to worry about further possible conditions
which the latter constraints might impose when just setting $H$ to
zero. Previously this was impossible as we recall now, as well.

We will show that $D_S W = 0 = D_T W$ can now have a solution
besides the runaway solution $s,t \ra \infty$ (the $D_i W=0$
conditions determine then the $z_i$). With $W^{S,T}=W_{WSI} +
W_{GC}$ one has
\beqa
D_S W^{S,T} &=& K_S \bigg( 2as \, W_{GC} + W^{S,T} \bigg)\\
D_T W^{S,T} &=& K_T \bigg(2\frac{bt}{3}\, W_{WSI} + 2 \frac{a\beta
t}{3}\, W_{GC} + W^{S,T}\bigg)
\eeqa
with the ensuing condition for a non-runaway solution
\beqa
\label{beta equation}
(2as+1)\Big(2\frac{bt}{3}+1\Big)=2 \frac{a\beta t}{3}+1
\eeqa

The point here is that this condition, as $a,s,b,t$ are positive,
only led to a contradiction in the previous case of $\beta=0$, but
no longer now. The system is now generically be solvable.

Note again that here the somewhat more non-standard choice $\beta
> 0$ would have to be made. This implies furthermore that
(\ref{beta no problem}) now gives\footnote{because one has
$3s-\beta t < 0$, as one can see from rewriting (\ref{beta
equation}) as $ 2asbt + a (3s-\beta t)+bt=0$.} $k\in {\bf R^{>0}}$
which would bring one back, with $\Re f_{obs} = -2s - \frac{b}{ak}
t$, to the transition problem.\footnote{The conclusions of this
one and the previous subsection will not change if one invokes the
correction $-\log \big( (T+\bar{T})^3 + E\big)$ with $E>0$ to the
K\"ahler potential in the $T$ sector.}

\vspace{.1cm}

\noindent {\em Remark 2: One-loop corrections which depend on the
complex structure moduli}

Actually, not only in the exponent of $W_{GC}=Ce^{-aS}$ a K\"ahler
modulus dependence should be included as a one-loop threshold
effect, but also a $z_i$-dependence of $C$ should be considered
(as was already done for $B$). As described in [\ref{ber}] for the
moduli-dependence of the threshold-corrections $\Delta$ the
Ray-Singer torsion will be relevant. More precisely if the
relevant $E_8$ is broken by a gauge bundle $V$ of structure group
$H$ to a (simple) group $G$ with ${\bf 248}=\oplus_k (R_k, r_k)$
with respect to $G \times H$ one finds (up to an additive piece
$\De(E_8)=30 RS({\bf C})$) $\De_G = \sum_k C_{R_k} RS(V_{r_k})=12
F_1$. Inclusion of this complex structure moduli dependence leads
to the replacement of (\ref{i cond}) by
\beqa
\label{prefactor zi dependence}
D_i W =  b_i \frac{vol}{i} + \frac{\del_i B}{B}W_{WSI} +
\frac{\del_i C}{C}W_{GC} + K_i ( W_{WSI} + W_{GC})
\eeqa
which means that, after imposing the demand $b_i=0$, (\ref{WST = 0
completed}) reads now
\beqa
\bigg(\frac{\del_i B}{BK_i}+1\bigg) W_{WSI} + \bigg(\frac{\del_i
C}{CK_i}+1\bigg) W_{GC}=0
\eeqa

\bigskip
\noindent {\large\bf Acknowledgements}\\[2ex]
We thank Bj\"orn Andreas, Melanie Becker, Gabriel Lopes Cardoso,
Peter Mayr, Susanne Reffert, Waldemar Schulgin and Stephan
Stieberger for discussion. We would also like to thank the referee
for a number of useful comments.
A.K.~has been supported by NSF grant PHY-0354401.

\section*{References}
\begin{enumerate}

\item
\label{KKLT}
S. Kachru, R. Kallosh, A. Linde and S.P. Trivedi,
{\em De Sitter Vacua in String Theory}, hep-th/0301240, Phys. Rev.
{\bf D68} (2003) 046005.

\item
\label{Rohm Witten}
R. Rohm and E. Witten,
{\em The Antisymmetric Tensor Field in Superstring Theory}, Ann.
of Phys. {\bf 170} (1986) 454.

\item
\label{BdA}
R. Brustein and S. P. de Alwis,
{\em Moduli Potentials in String Compactifications with Fluxes:
Mapping the Discretuum}, hep-th/0402088, Phys. Rev. {\bf D69}
(2004) 126006.

\item
\label{S}
A. Strominger,
{\em Superstrings with torsion}, Nucl. Phys. {\bf B274} (1986)
253.

\item
\label{CCdAL2}
G. L. Cardoso, G. Curio, G. Dall'Agata and D. L\"ust, {\em BPS Action
and Superpotential for Heterotic String Compactifications with
Fluxes}, hep-th/0306088, JHEP {\bf 0310} (2003) 004.

\item
\label{B1}
K. Becker, M. Becker, K. Dasgupta and P.S. Green,
{\em Compactifications of Heterotic Theory on Non-Kaehler Complex
Manifolds. {I}}, hep-th/0301161, JHEP {\bf 04} (2003) 007.

\item
\label{W}
E. Witten,
{\em World-Sheet Corrections Via D-Instantons}, hep-th/9907041,
JHEP {\bf 0002} (2000) 030.

\item
\label{BW}
C. Beasley and E. Witten,
{\em Residues and World-Sheet Instantons}, hep-th/0304115, JHEP
{\bf 0310} (2003) 065.

\item
\label{W QCD}
E. Witten,
{\em Branes And The Dynamics Of QCD}, hep-th/9706109, Nucl. Phys.
{\bf B507} (1997) 658.

\item
\label{Nilles}
K. Choi, A. Falkowski, H. P. Nilles, M. Olechowski and S. Pokorski,
{\em Stability of Flux Compactifications and the Pattern of
Supersymmetry Breaking}, hep-th/0411066, JHEP {\bf 0411} (2004)
076.

\item
\label{BKQ}
C.P. Burgess, R. Kallosh, F. Quevedo,
{\em De Sitter String Vacua from Supersymmetric D-terms},
hep-th/0309187, JHEP {\bf 0310} (2003) 056.

\item
\label{GKP}
S.B. Giddings, S. Kachru and J. Polchinski,
{\em Hierarchies from Fluxes in String Compactifications},
hep-th/0105097, Phys. Rev. {\bf D66} (2002) 106006.

\item
\label{GKLM}
S. Gukov, S. Kachru, X. Liu and L. McAllister,
{\em Moduli Stabilization with Fractional Chern-Simons
Invariants}, hep-th/0310159, Phys. Rev. {\bf D69} (2004) 086008.

\item
\label{fluxes}
S.~Gukov, C.~Vafa and E.~Witten,
{\em CFT's From Calabi-Yau Four-folds}, hep-th/9906070 Nucl.Phys.
{\bf B584} (2000) 69, Erratum-ibid. B608 (2001)
477;\\
S.~Gukov,
{\em Solitons, Superpotentials and Calibrations},
hep-th/9911011, Nucl.Phys. {\bf B574} (2000) 169;\\
T.~R.~Taylor and C.~Vafa,
{\em RR flux on Calabi-Yau and partial supersymmetry breaking},
Phys.\ Lett.\ B {\bf 474}, 130 (2000),
hep-th/9912152;\\
P.~Mayr,
{\em On supersymmetry breaking in string theory and its
realization in brane worlds}, Nucl.\ Phys.\ B {\bf 593}, 99
(2001),
hep-th/0003198;\\
G.~Curio, A.~Klemm, D.~L\"ust and S.~Theisen,
{\em On the vacuum structure of type II string compactifications
on  Calabi-Yau spaces with H-fluxes}, Nucl.\ Phys.\ B {\bf 609}, 3
(2001),
hep-th/0012213;\\
R.~Blumenhagen, D.~L\"ust and T.~R.~Taylor,
{\em Moduli stabilization in chiral type IIB orientifold models
with fluxes}, Nucl.\ Phys.\ B {\bf 663}, 319 (2003),
hep-th/0303016;\\
J.~F.~G.~Cascales and A.~M.~Uranga,
{\em Chiral 4d N = 1 string vacua with D-branes and NSNS and RR
fluxes}, JHEP {\bf 0305}, 011 (2003)
hep-th/0303024;\\
P.~G.~Camara, L.~E.~Ibanez and A.~M.~Uranga,
Nucl.\ Phys.\ B {\bf 689}, 195 (2004)
[arXiv:hep-th/0311241];\\
C.~Angelantonj, R.~D'Auria, S.~Ferrara and M.~Trigiante,
{\em $K3\times T^2/Z(2)$ orientifolds with fluxes, open string
moduli and critical points}, Phys.\ Lett.\ B {\bf 583}, 331 (2004)
[arXiv:hep-th/0312019];\\
M.~Grana, T.~W.~Grimm, H.~Jockers and J.~Louis,
{\em Soft supersymmetry breaking in Calabi-Yau orientifolds with
D-branes  and fluxes}, Nucl.\ Phys.\ B {\bf 690}, 21 (2004)
hep-th/0312232;\\
D.~L\"ust, S.~Reffert and S.~Stieberger,
{\em Flux-induced soft supersymmetry breaking in chiral type IIb
orientifolds with D3/D7-branes}, Nucl.\ Phys.\ B {\bf 706}, 3
(2005)
hep-th/0406092;\\
F.~Marchesano and G.~Shiu,
{\em MSSM vacua from flux compactifications}, Phys.\ Rev.\ D {\bf
71}, 011701 (2005)
hep-th/0408059;\\
M.~Cvetic and T.~Liu,
{\em Supersymmetric Standard Models, Flux Compactification and
Moduli Stabilization},
hep-th/0409032;\\
L.~G\"orlich, S.~Kachru, P.~K.~Tripathy and S.~P.~Trivedi,
{\em Gaugino condensation and nonperturbative superpotentials in
flux compactifications},
hep-th/0407130;\\
D.~L\"ust, P.~Mayr, S.~Reffert and S.~Stieberger,
{\em F-theory flux, destabilization of orientifolds and soft terms
on D7-branes},
hep-th/0501139;\\
H.~Jockers and J.~Louis,
{\em D-terms and F-terms from D7-brane fluxes}, hep-th/0502059.

\item
\label{Green}
G.W. Gibbons, M.B. Green and M.J. Perry,
{\em Instantons and seven-branes in type IIB superstring theory},
hep-th/9511080, Phys.Lett. {\bf B370} (1996) 37.

\item
\label{Schulgin}
W. Schulgin et al., work in progess.

\item
\label{D}
M.~R.~Douglas,
{\em The statistics of string / M theory vacua}, JHEP {\bf 0305},
046 (2003)
[arXiv:hep-th/0303194;\\
F. Denef and M.R. Douglas,
{\em Distributions of Flux Vacua},
hep-th/0404116, JHEP {\bf 0405} (2004) 072.\\
A. Giryavets, S. Kachru and P.K. Tripathy,
{\em On the Taxonomy of Flux Vacua},
hep-th/0404243, JHEP {\bf 0408} (2004) 002.\\
J.P. Conlon and F. Quevedo,
{\em On the Explicit Construction and Statistics of Calabi-Yau
Flux Vacua}, hep-th/0409215, JHEP {\bf 0410} (2004) 039.

\item \label{Dine}
M. Dine, D. O'Neil, Z. Sun,
{\em Branches of the Landscape}, hep-th/0501214.

\item
\label{DRSW}
S. Ferrara, L. Girardello and H.P. Nilles,
{\em Breakdown Of Local Supersymmetry Through Gauge Fermion
Condensates},
Phys. Lett. {\bf B125} (1983) 457;\\
M. Dine, R. Rohm, N. Seiberg and E. Witten,
{\em Gluino Condensation in Superstring Models},
Phys. Lett. {\bf B156} (1985) 55;\\
J.P. Derendinger, L. E. Ibanez and H. P. Nilles,
{\em On The Low-Energy D = 4, N=1 Supergravity Theory Extracted
>From The D = 10, N=1 Superstring}, Phys. Lett. {\bf B155} (1985)
65;\\
J.P. Derendinger, L.E. Ibanez and H. P. Nilles,
{\em On The Low-Energy Limit Of Superstring Theories},
Nucl. Phys. {\bf B267} (1986) 365;\\
A. Font, L.E. Ibanez, D. L\"ust and F. Quevedo,
{\em Supersymmetry Breaking From Duality Invariant Gaugino
Condensation},
Phys. Lett. {\bf B245} (1990) 401;\\
S. Ferrara, N. Magnoli, T.R. Taylor and G. Veneziano,
{\em Duality And Supersymmetry Breaking In String Theory},
Phys. Lett. {\bf B245} (1990) 409;\\
H.P. Nilles and M. Olechowski,
{\em Gaugino Condensation And Duality Invariance}, Phys. Lett.
{\bf B248} (1990) 268.

\item
\label{CCdAL1}
G. L. Cardoso, G. Curio, G. Dall'Agata, D. L\"ust, P. Manousselis and
G. Zoupanos, {\em Non-K\"ahler String Backgrounds and their Five Torsion
Classes}, hep-th/0211118, Nucl. Phys. {\bf B652} (2003) 5.

\item \label{Gauntlett1}
J.P. Gauntlett, D. Martelli and D. Waldram,
{\em Superstrings with Intrinsic Torsion}, hep-th/0302158.

\item \label{Gauntlett2}
J.P. Gauntlett, N.W. Kim, D. Martelli and D. Waldram,
{\em Fivebranes Wrapped on SLAG Three-Cycles and Related
Geometry,} hep-th/0110034, JHEP {\bf 0111} (2001) 018.

\item \label{B2}
K. Becker, M. Becker, K. Dasgupta and S. Prokushkin,
{\em Properties of Heterotic Vacua from Superpotentials},
hep-th/0304001, Nucl. Phys. {\bf B666} (2003) 144.

\item
\label{CCdAL3}
G. L. Cardoso, G. Curio, G. Dall'Agata and D. L\"ust,
{\em Heterotic String Theory on non-Kaehler Manifolds with H-Flux
and Gaugino Condensate}, hep-th/0310021, Fortsch.Phys. {\bf 52}
(2004) 483.

\item
\label{CCdAL4}
G. L. Cardoso, G. Curio, G. Dall'Agata and D. L\"ust,
{\em Gaugino Condensation and Generation of Supersymmetric 3-Form
Flux}, hep-th/0406118, JHEP {\bf 0409} (2004) 059.

\item \label{BBCQ}
V. Balasubramanian, P. Berglund, J. Conlon and F. Quevedo,
{\em Systematics of Moduli Stabilisation in Calabi-Yau Flux
Compactifications}, hep-th/0502058.

\item
\label{NN}
A. Niemeyer and H.P. Nilles,
{\em Gaugino Condensation and the Vacuum Expectation Value of the
Dilaton}, hep-th/9508173.

\item
\label{X}
R. Xiu,
{\em Supersymmetry Breaking Scheme and The Derivation of
$M_{GUT}=10^{16}GeV$ from A String Model}, hep-ph/9412262.

\item
\label{CK2}
G. Curio and A. Krause,
{\em G-Fluxes and Non-Perturbative Stabilization of Heterotic
M-Theory}, hep-th/0108220, Nucl. Phys. {\bf B643} (2002) 131.

\item
\label{BCK}
M. Becker, G. Curio and A. Krause,
{\em De Sitter Vacua from Heterotic M-Theory}, hep-th/0403027,
Nucl. Phys. {\bf B693} (2004) 223.

\item
\label{BO}
E.I. Buchbinder and B.A. Ovrut,
{\em Vacuum Stability in Heterotic M-Theory}, hep-th/0310112,
Phys. Rev. {\bf D69} (2004) 086010.

\item
\label{Witten 86}
E. Witten,
{\em New Issues of Manifolds of SU(3) Holonomy},
Nucl. Phys. {\bf B268} (1986) 79.\\
M.B. Green, J.H. Schwarz and
E. Witten, {\em Superstring Theory}, vol II, Cambridge University Press
(1987). \\
X. Wu and E. Witten, {\em Space-Time Supersymmetry in Large Radius
Expansion of Superstring Compactification}, Nucl. Phys. {\bf B
289} (1987) 385.

\item
\label{HetMWarp}
E. Witten,
{\em Strong Coupling Expansion of Calabi-Yau Compactification},
hep-th/9602070, Nucl. Phys. {\bf B471} (1996) 135.\\
G. Curio and A. Krause,
{\em Four-Flux and Warped Heterotic M-Theory Compactifications},
hep-th/0012152, Nucl. Phys. {\bf B602} (2001) 172.\\
G. Curio and A. Krause,
{\em Enlarging the Parameter Space of Heterotic M-Theory Flux
Compactifications to Phenomenological Viability}, hep-th/0308202,
Nucl. Phys. {\bf B693} (2004) 195.

\item
\label{FKLZ}
V. Kaplunovsky, {\em One-Loop Threshold-Effects in String Unification},
hep-th/9205070, Nucl. Phys. {\bf B307} (1988) 145; Erratum hep-th/
9205068, {\bf B382} (1992) 436.\\
L.J. Dixon, V. Kaplunovsky and J. Louis,
{\em Moduli Dependence of String Loop Corrections to Gauge
Coupling Constants},
Nucl. Phys. {\bf B355} (1991) 649.\\
S. Ferrara, C. Kounnas, D. L\"ust and F. Zwirner,
{\em Duality Invariant Partition Functions and Automorphic
Superpotentials for (2,2) String Compactifications},
Nucl. Phys. {\bf B365} (1991) 431.\\
J.P. Derendinger, S. Ferrara, C. Kounnas and F. Zwirner,
{\em One-Loop Corrections to String Effective Field Theories:
Field Dependent Gauge Couplings and Sigma Model Anomalies},
Nucl. Phys. {\bf B372} (1992) 145.\\
I. Antoniadis, E. Gava and K.S. Narain,
{\em Moduli Corrections to Gauge and Gravitational Couplings in
Four Dimensional Superstrings},
hep-th/9204030, Nucl. Phys. {\bf B383} (1992) 93 .\\
I. Antoniadis, E. Gava and K.S. Narain,
{\em Moduli Corrections to Gravitational Couplings from String
Loops}, hep-th/9203071 , Phys. Lett. {\bf B283} (1992) 209.\\
H.~P.~Nilles and S.~Stieberger,
{\em String unification, universal one-loop corrections and strongly coupled
heterotic string theory},
Nucl.\ Phys.\ B {\bf 499}, 3 (1997),
hep-th/9702110.\\
S.~Stieberger,
{\em (0,2) heterotic gauge couplings and their M-theory origin},
Nucl.\ Phys.\ B {\bf 541}, 109 (1999),
hep-th/9807124.

\item
\label{ber}
M. Bershadsky, S. Cecotti, H. Ooguri and C. Vafa,
{\em Kodaira-Spencer Theory of Gravity and Exact Results for
Quantum String Amplitudes}, hep-th/9309140, Commun. Math. Phys.
165 (1994) 311.

\end{enumerate}

\end{document}